\documentclass[12pt]{iopart}
\usepackage{iopams}
\usepackage{epsfig}
\usepackage{textcomp}
\usepackage{citesort}
\begin{document}

\title[Evolutionary advantages of adaptive rewarding]{Evolutionary advantages of adaptive rewarding}

\author{Attila Szolnoki$^1$ and Matja{\v z} Perc$^2$}
\address{$^1$Institute of Technical Physics and Materials Science, Research Centre for Natural Sciences, Hungarian Academy of Sciences, P.O. Box 49, H-1525 Budapest, Hungary\\
$^2$Department of Physics, Faculty of Natural Sciences and Mathematics, University of Maribor, Koro{\v s}ka  cesta 160, SI-2000 Maribor, Slovenia}
\ead{szolnoki.attila@ttk.mta.hu, matjaz.perc@uni-mb.si}

\begin{abstract}
Our wellbeing depends as much on our personal success, as it does on the success of our society. The realization of this fact makes cooperation a very much needed trait. Experiments have shown that rewards can elevate our readiness to cooperate, but since giving a reward inevitably entails paying a cost for it, the emergence and stability of such behavior remain elusive. Here we show that allowing for the act of rewarding to self-organize in dependence on the success of cooperation creates several evolutionary advantages that instill new ways through which collaborative efforts are promoted. Ranging from indirect territorial battle to the spontaneous emergence and destruction of coexistence, phase diagrams and the underlying spatial patterns reveal fascinatingly reach social dynamics that explains why this costly behavior has evolved and persevered. Comparisons with adaptive punishment, however, uncover an Achilles heel of adaptive rewarding that is due to over-aggression, which in turn hinders optimal utilization of network reciprocity. This may explain why, despite of its success, rewarding is not as firmly weaved into our societal organization as punishment.
\end{abstract}

\pacs{87.23.Ge, 89.75.Fb, 89.65.-s}
\maketitle

\section{Introduction}
\label{intro}
Responsible usage of public goods and continuous investments into the common pool are of paramount importance for sustainable development on a global scale. Loosing sight of this by over-exploiting the goods for short-term benefits inevitably creates systemic risks that may lead to the ``tragedy of the commons'' \cite{hardin_g_s68}. The public goods game captures succinctly the essence of the underlying social dilemma by requiring that players decide simultaneously whether they wish to bare the cost of cooperation and thus to contribute to the common pool, or not. Regardless of their decision, each member of the group receives an equal share of the public good after the initial contributions are multiplied by a factor that takes into account the added value of collaborative efforts. Individuals are best off by defecting, while the group is most successful if everybody cooperates. Historical evidence suggest that humans have developed remarkable other-regarding abilities to mitigate between-group conflicts \cite{bowles_11}, as well as to help each other by rearing offspring that survived \cite{hrdy_11}. However, while these issues might have sparked our cooperative behavior, it is mechanisms like kin and group selection as well as different forms of reciprocity \cite{nowak_s06} or other recently identified mechanisms \cite{pinheiro_njp12,fort_jsm05,yang_hx_epl12}, that have likely been instrumental in eliciting its full potential and solidifying it as one of the most distinguishable behavioral traits of the mankind.

Reward and punishment \cite{sigmund_pnas01} are also cited frequently as viable means to promote the evolution of public cooperation, although punishment has received substantially more attention, as reviewed comprehensively in \cite{sigmund_tee07}. Related to that, it is important to note that recent research related to antisocial punishment \cite{herrmann_s08,rand_nc11} and reward in particular \cite{rand_s09}, is questioning the aptness of sanctioning for elevating collaborative efforts and raising social welfare. Indeed, while the majority of previous studies addressing the ``stick versus carrot'' dilemma concluded that punishment is more effective than reward in sustaining cooperation \cite{sigmund_pnas01,sigmund_tee07}, evidence suggesting that rewards may be as effective as punishment and lead to higher total earnings without potential damage to reputation \cite{milinski_n02} or fear from retaliation \cite{dreber_n08} is mounting rapidly. Moreover, in their recent paper \cite{rand_nc11}, Rand and Nowak provide firm evidence that antisocial punishment renders the concept of sanctioning ineffective, and argue further that healthy levels of cooperation are likelier to be achieved through less destructive means.

Regardless of whether we place the burden of cooperation promotion on punishment \cite{hauert_s07,gachter_s08,rockenbach_n09,boyd_s10} or reward \cite{hilbe_prsb10,szolnoki_epl10,hauert_jtb10}, the problem with both actions is that they are costly. In particular, punishment implies paying a cost for another person to incur a cost, while rewards obviously incorporate a cost to bear too, but for another person to experience a benefit. Cooperators who abstain therefore become ``second-order free-riders'', and they can seriously challenge the success of sanctioning \cite{panchanathan_n04,fehr_n04,fowler_n05b} as well as rewarding \cite{szolnoki_epl10}. Here we focus on the later and take into account the fact that our willingness to reward others depends sensitively on the success of antisocial behavior. If defection is on the rise, we may feel more inclined to support cooperation by means of additional incentives in order to avert an impending social decline. On the other hand, if everybody is already cooperating such actions may appear superfluous. Moreover, there is a permanent tendency to eschew the costs that are associated with administrating rewards. Inspired by these observations, we introduce a third strategy to the spatial public goods game to supplement the traditional cooperators and defectors, namely the so-called rewarding cooperators, and show that adaptive rewarding yields several evolutionary advantages that can overcome the ``second-order free-rider'' problem. Compared to steady rewarding \cite{szolnoki_epl10}, for example, the cyclic dominance between the three competing strategies can be broken, which in turn leads to higher levels of cooperation and even to completely defector-free states. Punishment, nevertheless, still outperforms rewarding for it acts more coherently with network reciprocity. We thus arrive at interesting and partly counterintuitive conclusions that extend the existing theory on sanctioning and rewarding in structured populations \cite{brandt_prsb03,nakamaru_eer05,helbing_ploscb10,szolnoki_epl10,perc_njp12}, as well as supplement the array of recently identified mechanisms that promote cooperation in public goods games, ranging from complex interaction networks and coevolution \cite{lozano_ploso08,wu_t_epl09,wu_t_pre09,gomez-gardenes_c11,gomez-gardenes_epl11} over diversity \cite{santos_n08,fort_epl08,wang_j_pre10b,perc_njp11,santos_jtb12} to the risk of collective failures \cite{santos_pnas11} and selection pressure \cite{van-segbroeck_njp11}. Before presenting the main results, however, we proceed with a detailed description of the studied spatial public goods game.

\section{Spatial public goods game with adaptive rewarding}
\label{Model}
The game is contested by cooperators ($s_x = C$), defectors ($s_x = D$) and rewarding cooperators ($s_x = R$), who initially populate the square lattice with equal probability. A player $x$ plays the public goods game with its $k=G-1=4$ interaction partners as a member of all the $g=1, \ldots, G=5$ groups it belongs to. Both cooperating strategies contribute $1$ to the public good while defectors contribute nothing. The sum of all contributions in each group is multiplied by the factor $r>1$, reflecting the synergetic effects of cooperation, and the resulting amount is equally divided amongst all group members irrespective of their strategy. Adaptive rewarding is accommodated by assigning each rewarding cooperator an additional parameter $\pi_x$, which keeps score of the rewarding activity. While this parameter is initially zero, subsequently, whenever a defector succeeds in passing its strategy, all the remaining rewarding cooperators in all the groups containing the defeated player increase their rewarding activity by one, i.e., $\pi_x=\pi_x+1$. The related costs increase accordingly. However, to maintain the latter is unwanted, and hence at every second round all rewarding cooperators decrease their rewarding activity by one, as long as $\pi_x \geq 0$. The payoff of player $x$ adopting $s_x = C$ in a given group $g$ of size $G$ is thus
\begin{equation}
P_C^g=r \frac{N_C+N_R+1}{G} - 1 + \frac{\Delta}{k} \sum_{i \in g}\pi_i  \,\, ,
\label{PC}
\end{equation}
where $N_C$, $N_D$ and $N_R$ are the number of other cooperators, defectors and rewarding cooperators in the group $g$, respectively. The sum runs across all the neighbors in the group, while $\pi_i$ is the actual rewarding activity of player $i$. The corresponding payoff of a rewarding cooperator at site $x$ is
\begin{equation}
P_R^g=P_C^g - \frac{\alpha \Delta}{k} \pi_x (N_C+N_R) \,\, ,
\label{PR}
\end{equation}
while a defector, who's payoff is derived exclusively from the contributions of others, gets
\begin{equation}
P_D^g=r \frac{N_C+N_R}{G} \,\, .
\label{PD}
\end{equation}
As it follows, each player adopting $s_x = C$ or $s_x = R$ is rewarded with an amount $\pi_i \Delta/k$ from every rewarding cooperator, having rewarding activity $\pi_i$, that is a member of the same group. At the same time, each rewarding cooperator bares the cost $\pi_i \alpha \Delta/k$ for every cooperator that was rewarded. Self-rewarding is excluded. Here $\Delta$ and $\alpha$ are important free parameters, determining the incremental step used for the rewarding activity and the cost of rewards, respectively. Note that $\alpha$ is actually the ratio between the cost of rewarding and the reward that is allotted to cooperators.

The stationary fractions of cooperators $\rho_C$, defectors $\rho_D$ and rewarding cooperators $\rho_R$ on the square lattice are determined by means of a random sequential update comprising the following elementary steps. First, a randomly selected player $x$ plays the public goods game with its partners as a member of all the five groups it belongs to. The overall payoff it thereby obtains is thus $P_{s_x} = \sum_g P_{s_x}^g$. Next, one of the four nearest neighbors of player $x$ is chosen randomly. This player $y$ also acquires its payoff $P_{s_y}$ identically as previously player $x$. Finally, if $s_x \neq s_y$ player $y$ imitates the strategy of player $x$ with the probability $q=1/\{1+\exp[(P_{s_y}-P_{s_x})/K]\}$, where $K$ determines the level of uncertainty by strategy adoptions. Without loss of generality we set $K=0.5$ \cite{szolnoki_pre09c}, implying that better performing players are readily imitated, but it is not impossible to adopt the strategy of a player performing worse. Each full Monte Carlo step of the game involves all players having a chance to adopt a strategy from one of their neighbors once on average. Depending on the proximity to phase transition points and the typical size of emerging spatial patterns, the linear system size was varied from $L=200$ to $2000$ and the equilibration required up to $10^6$ full rounds of the game for the finite size effects to be avoided.

It is worth noting that this set-up enables us to directly compare the effectiveness of adaptive rewarding with steady rewarding efforts studied previously in \cite{szolnoki_epl10}. While the simulation details are identical in both cases, in the steady rewarding model players adopting
$s_x = R$ always reward every cooperator with a reward $\Delta/k$ and therefore bare the cost of rewarding $\alpha \Delta /k$. The initially set rewarding activity of rewarding cooperators $\pi_x=1$ never increases or decreases, while $\Delta$ simply determines the strength of rewards. As in the adaptive model, $\alpha$ determines just how costly rewards are. For further details we refer to \cite{szolnoki_epl10}, where the steady rewarding model was presented and studied in detail. Moreover, the outcome of the presently studied model can also be compared to the one obtained by means of adaptive punishment, as studied recently in \cite{perc_njp12}. The main difference is that while rewarding cooperators increase their rewarding activity to reward cooperators, punishing cooperators increase their punishing activity to punish defectors. In both cases a constant drift towards inactivity in terms of either punishment or reward is assumed. For further details we again refer to \cite{perc_njp12}, while here we proceed with presenting the main results.

\section{Results}
\label{results}

\subsection{Adaptive versus steady rewarding}
\label{expected}

\begin{figure}
\centerline{\epsfig{file=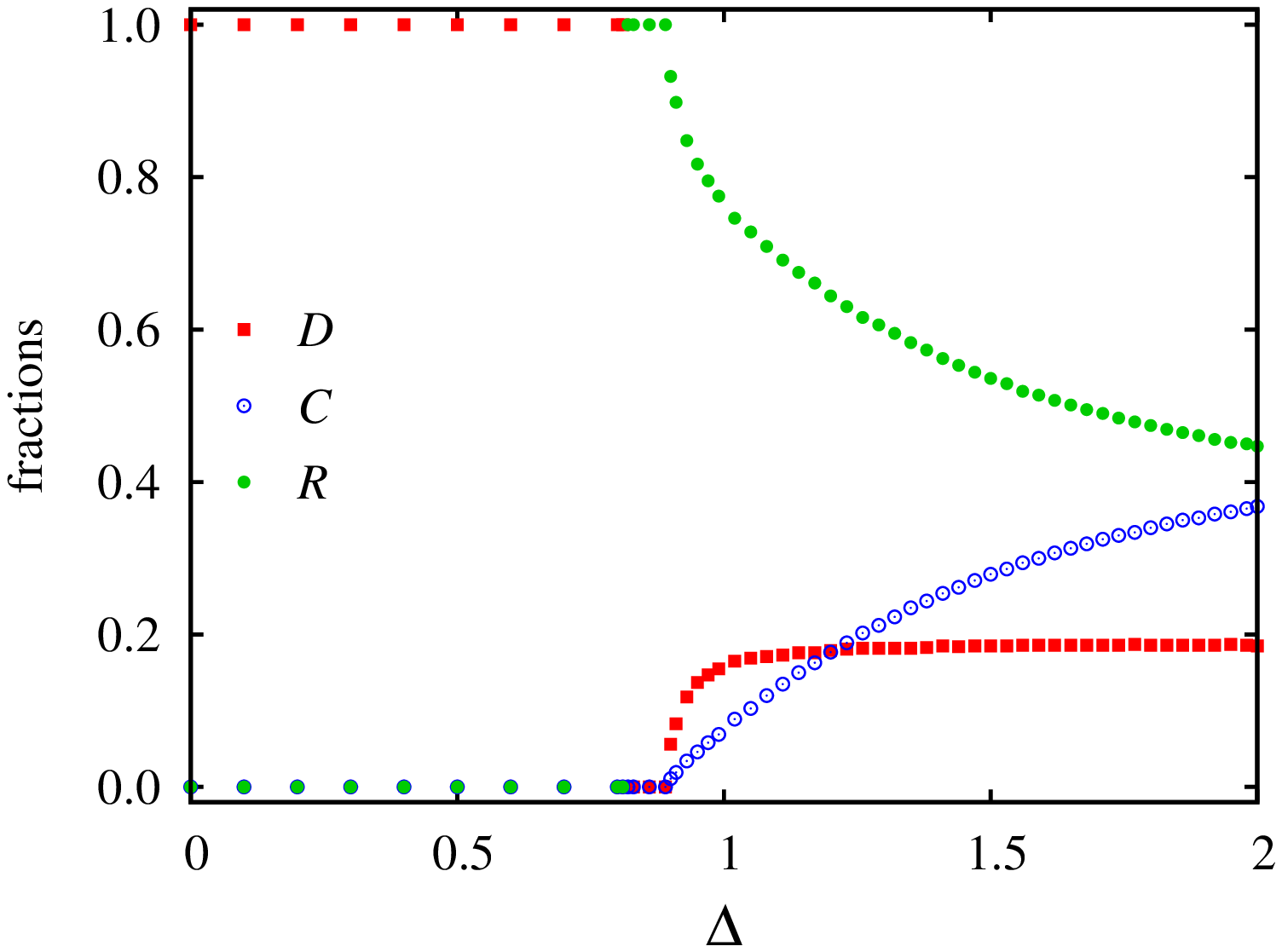,width=8cm}\epsfig{file=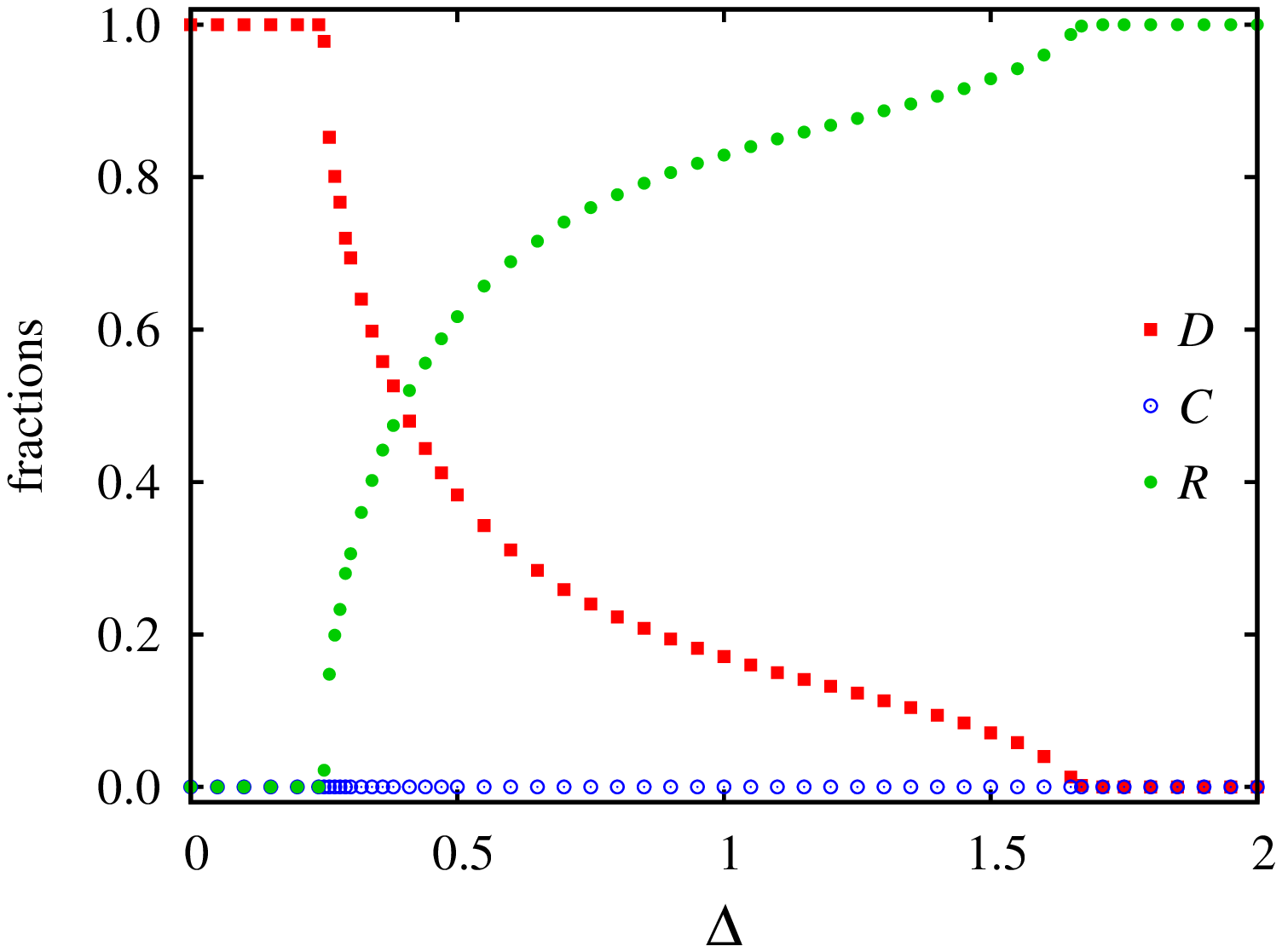,width=8cm}}
\caption{Fractions of the three competing strategies in dependence on $\Delta$, as obtained at $r=2$ and $\alpha=0.1$ for steady (left) and adaptive (right) rewarding. While steady rewarding fails to eliminate defection due to the spontaneous emergence of cycling dominance that is brought about by ``second-order free-riding'', adaptive rewarding suffers from no such drawbacks, gradually leading to complete dominance of rewarding cooperators as $\Delta$ increases.}
\label{motivate}
\end{figure}

Firstly, it is instructive to compare the impact of the newly introduced adaptive rewarding with that of steady rewarding at the same synergy factor $r$ and the cost of reward $\alpha$. As shown in Fig.~\ref{motivate} (left), the application of steady rewards yields a stable presence of defectors virtually across the whole span of $\Delta$. This implies that no matter how strong the rewarding, defection cannot be eliminated. Here rewarding cooperators enable the survival of cooperators, which act as ``second-order free-riders'', who in turn provide easy targets for defectors, thus creating a closed loop of dominance. The persistence of defectors is thus a direct consequence of ``second-order free-riding'', which emerges almost as soon as rewarding cooperators are able to invade defectors. Notably though, there is a very narrow span of intermediate $\Delta$ values, at which steady rewarding is just successful enough to overcome defection, but not sufficiently so to enable cooperators to free-ride on the newly acquired success. For adaptive rewarding, however, the outcome is significantly different, as shown in Fig.~\ref{motivate} (right). To begin with, much lower values of $\Delta$ suffice to elicit the downfall of defectors. But even more importantly, ``second-order free-riding'' never gets a foothold in the population. Accordingly, as $\Delta$ increases rewarding cooperators gradually rise to complete dominance, despite of the very low synergy factor ($r=2$) governing the production of public goods. As demonstrated in \cite{szolnoki_epl10}, defector-free states are attainable also with steady rewarding, but require $\alpha < 0.05$, i.e., very low costs of administrating the rewards. Adaptive rewarding is thus more effective, predominantly because ``second-order free-riders'' fail to induce cycling dominance between the three competing strategies.

\subsection{Phase diagrams and spatial patterns}
\label{phd}

\begin{figure}
\centerline{\epsfig{file=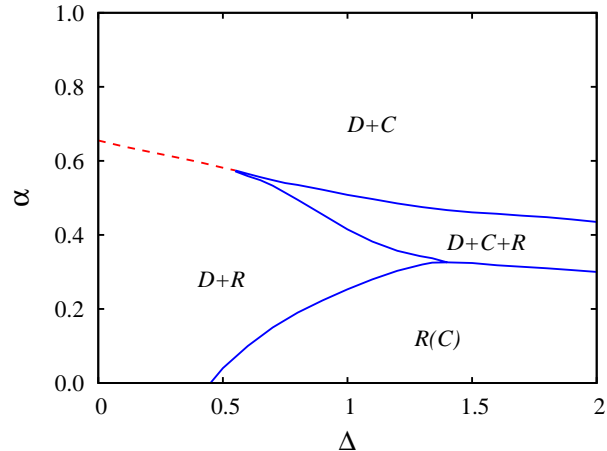,width=8cm}}
\caption{Full $\Delta-\alpha$ phase diagram, as obtained at $r=4.4$. Red dashed line depicts first-order phase transitions while blue solid lines depict continuous second-order phase transitions. Symbols mark the surviving strategies in the stationary state. Besides stable two strategy $D+C$ and $D+R$ phases, the coexistence of all three competing strategies is also possible, where $D$ and $C$ form an alliance to compete against $R$. Notably, $R(C)$ denotes the defection-free phase, but since in the absence of defectors strategies $R$ and $C$ become equivalent, the evolutionary process proceeds via slow logarithmic coarsening, as in the voter model \cite{dornic_prl01}. However, since at the time of extinction of defectors the majority of players are rewarding cooperators, the system finally arrives at the $R$ phase with a significantly higher probability. Notably, the dominance of strategy $R$ becomes more evident if rare mutations are allowed, similarly as reported for punishment in \cite{helbing_pre10c}.}
\label{phd_r4_4}
\end{figure}

The comparison with steady rewarding begets further explorations. In particular, the question is whether coexistence in the absence of cyclic dominance is nevertheless possible, and to what degree the results presented in Fig.~\ref{motivate} are robust to parameter variations? To address this systematically, we proceed with the presentation of characteristic phase diagrams and spatial patterns for different values of $r$.

\begin{figure}
\centerline{\epsfig{file=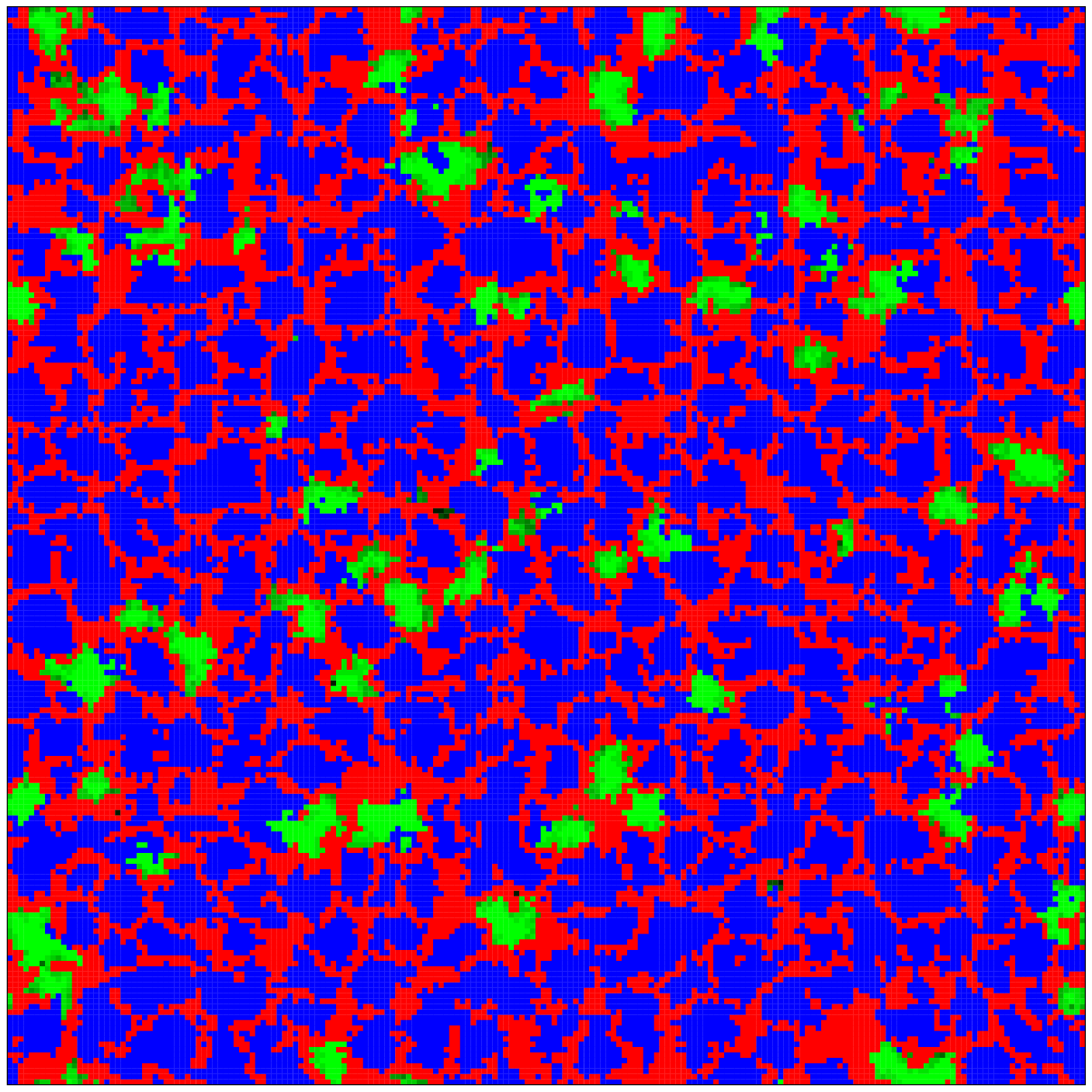,width=5cm}\epsfig{file=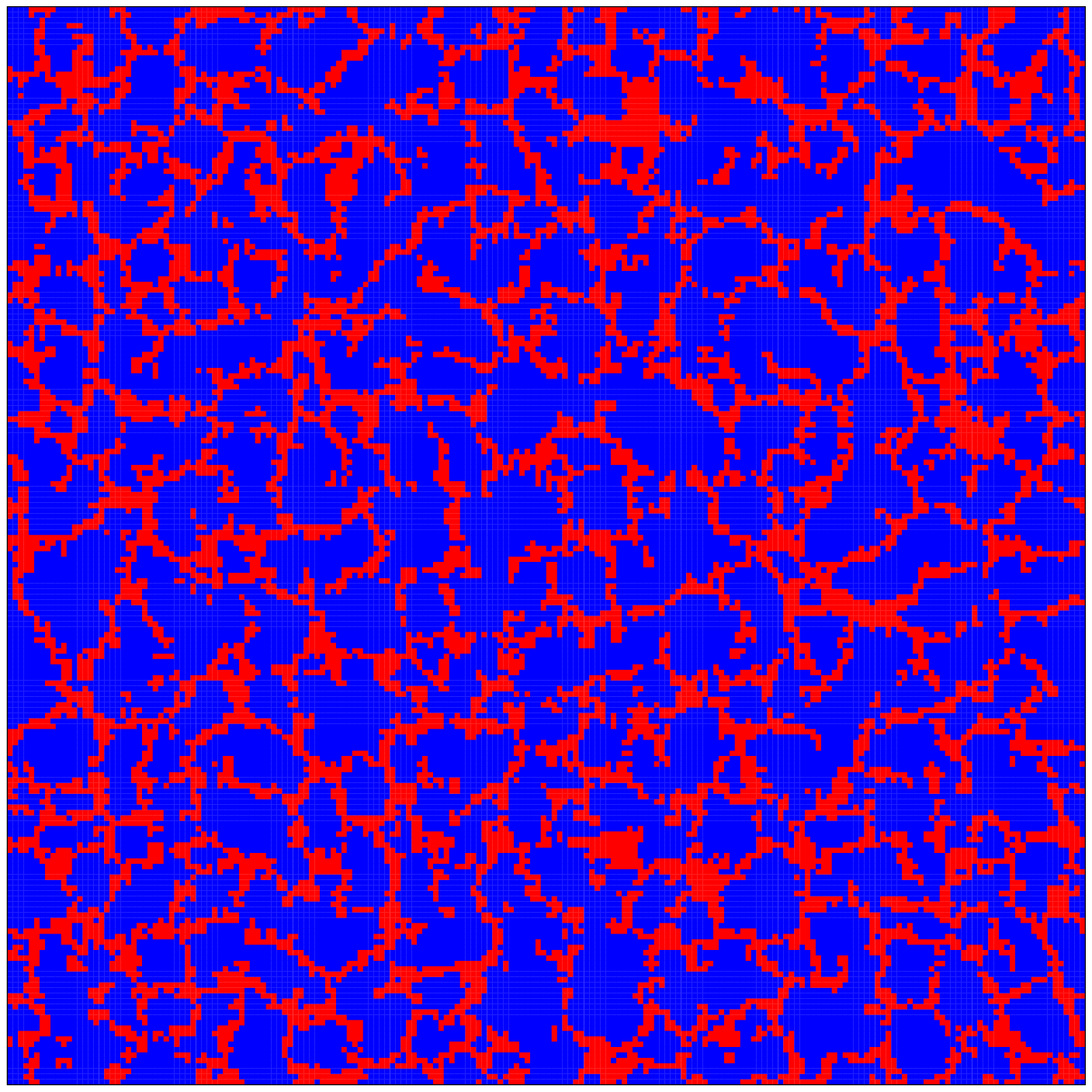,width=5cm}}
\centerline{\epsfig{file=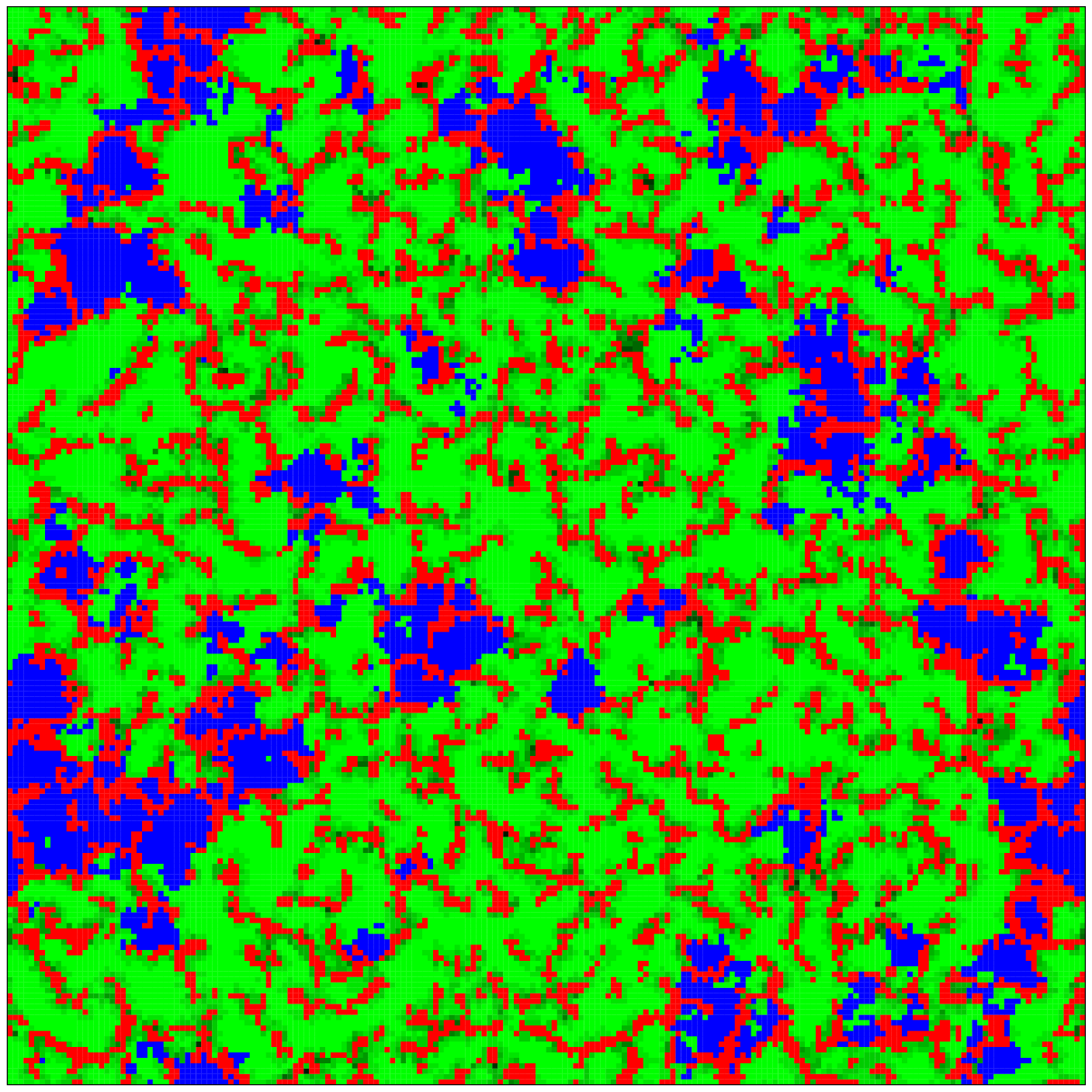,width=5cm}\epsfig{file=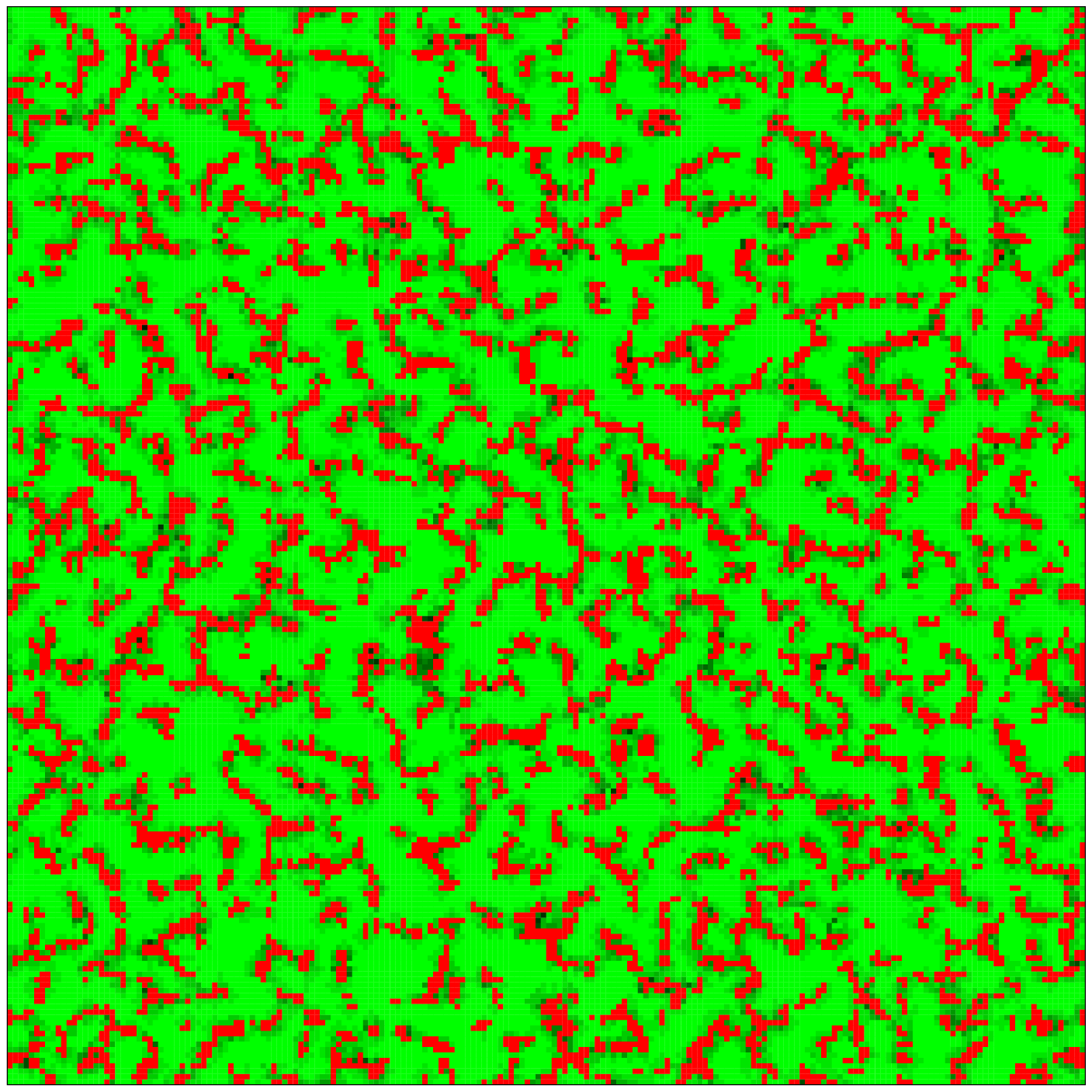,width=5cm}}
\caption{Indirect territorial competition between cooperators $C$ (blue) and rewarding cooperators $R$ (green) that is mediated by defectors $D$ (red). In the upper row cooperators are more successful in invading defectors. Accordingly, rewarding cooperators are crowded out, leaving a stable $D+C$ phase (top right) as the final stationary state. In the bottom row the situation is reversed. Rewarding cooperators outperform cooperators in the indirect competition against defectors, ultimately arriving at a stable $D+R$ phase. The spatial segregation of indirectly competing strategies against a third strategy (defectors) creates the blueprint for a discontinuous phase transition, as marked by the red dashed $D+C \to D+R$ transition line in Fig.~\ref{phd_r4_4}. Parameter values are: $r=4.4$, $\Delta=0.4$, $\alpha=0.9$ (top row) and $\alpha=0.5$ (bottom row).}
\label{indirect}
\end{figure}

Figure~\ref{phd_r4_4} features the full $\Delta-\alpha$ phase diagram for $r=4.4$. It is important to note that for such a relatively high value of $r$ cooperators can survive in the presence of defectors without rewards, solely on the basis of spatial reciprocity. Accordingly, if rewarding is inefficient and costly, rewarding cooperators die out, leaving $D+C$ as the stable two-strategy phase. As $\alpha$ decreases, however, rewarding cooperators become more and more competitive, which culminates in the outbreak of the stable $D+R$ phase if $\Delta$ is sufficiently small. Yet the discontinuous $D+C \to D+R$ transition is deceiving, in that it suggests that the competition is won or lost directly between cooperators and rewarding cooperators. This is in fact not the case because in the absence of defectors the relation between the two eventually becomes neutral. The victor between $C$ and $R$ is therefore determined indirectly in terms of which of the two strategies is more successful in invading defectors. This indirect territorial battle is illustrated in Fig.~\ref{indirect}, where in the upper row cooperators are more successful, while in the bottom row rewarding cooperators prevail. Note that in both cases cooperators and rewarding cooperators form compact clusters that are isolated from one another, which is a direct consequence of coarsening within a finite size domain. An identical phenomenon was reported in \cite{helbing_ploscb10}, where punishing cooperators and cooperators (``second-order free-riders'') engaged in indirect competition that was mediated by defectors, and where too the victor was determined based on the success and efficiency of this invasion. It is also worth emphasizing that the fraction of defectors changes insignificantly during this evolutionary process, regardless of whether finally the $D+C$ or the $D+R$ phase is reached, i.e., $C(R)$ spread almost exclusively on the expense of R(C) (not shown). Thus, defectors truly just mediate the difference in efficiency between cooperators and rewarding cooperators.

\begin{figure}
\centerline{\epsfig{file=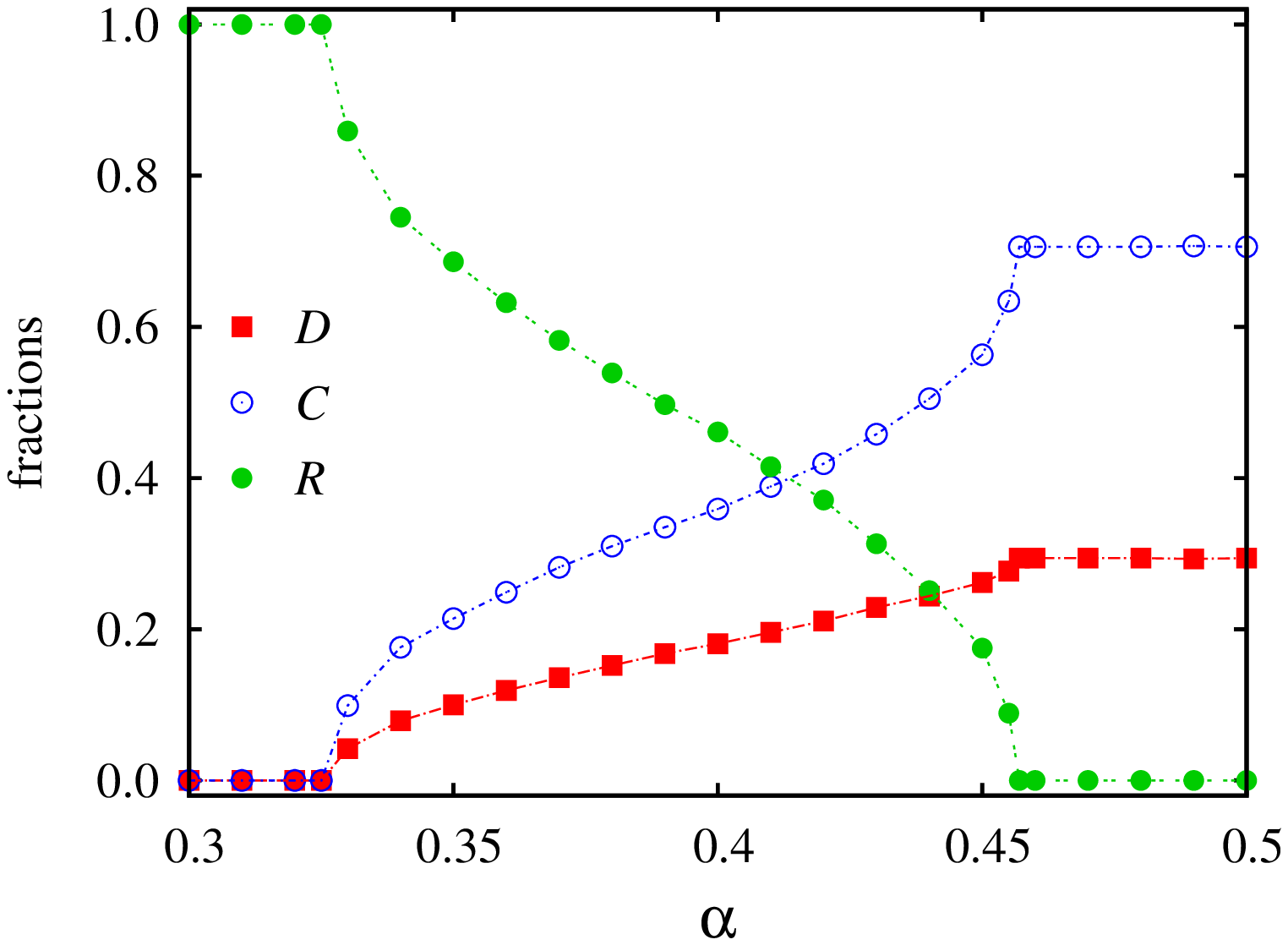,width=8cm}\epsfig{file=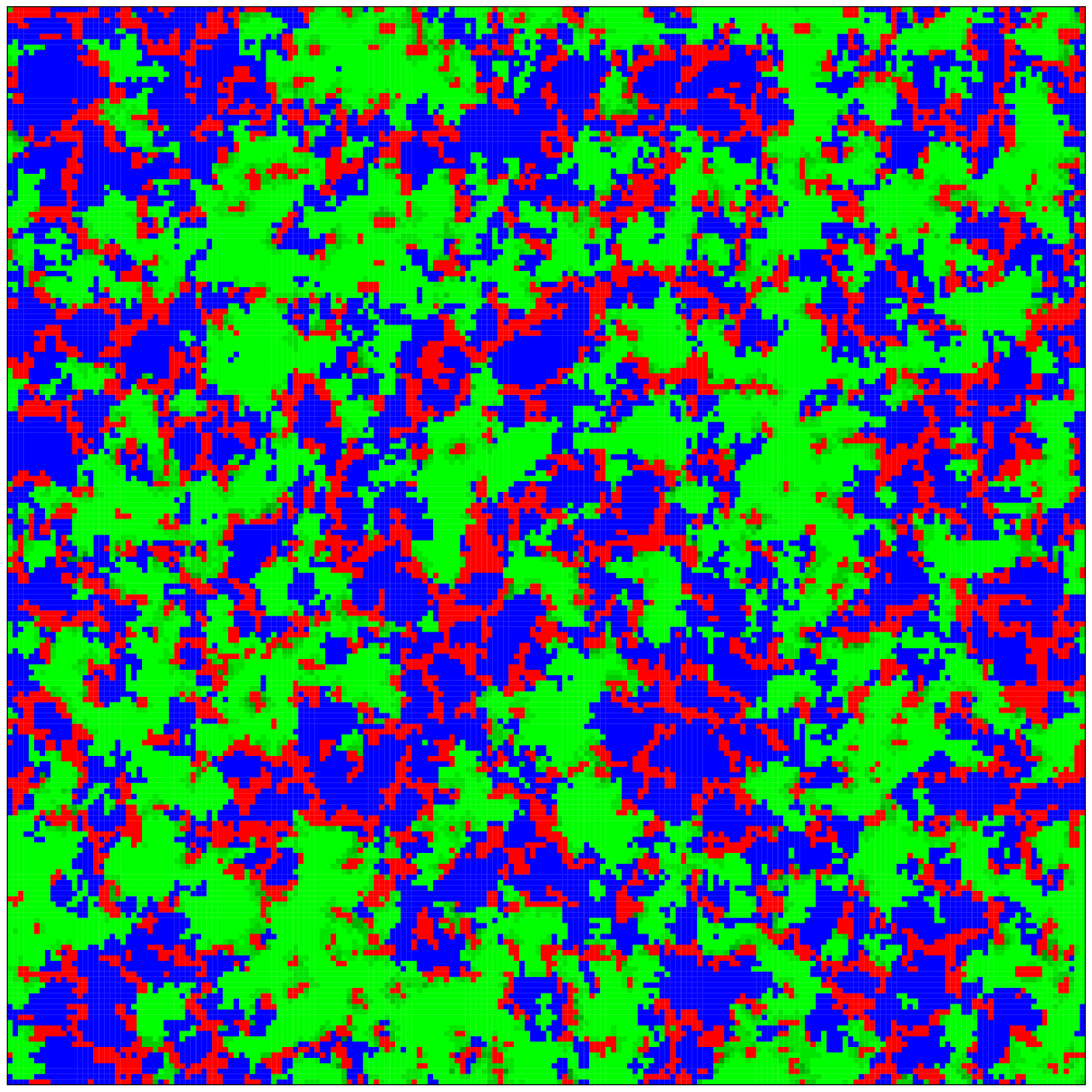,width=6cm}}
\caption{Left panel features a cross section of the phase diagram presented in Fig.~\ref{phd_r4_4}, as obtained at $\Delta=1.5$. As $\alpha$ increases the rewarding cooperators first give way to a three-strategy $D+C+R$ phase, while further on they completely subdue to the free-riding $D+C$ alliance. Right panel depicts a characteristic snapshot of the three-strategy phase taken at $\Delta=1.4$ and $\alpha=0.42$, where the $D+C$ alliance (red and blue) competes against invading $R$ (green). Note that here the $D+C$ patches are practically identical with the stationary state depicted in the top right panel of Fig.~\ref{indirect}.}
\label{alliance}
\end{figure}

Returning to the phase diagram presented in Fig.~\ref{phd_r4_4}, it can be observed that as $\Delta$ increases, the discontinuous first-order phase transitions give way to a continuous transition line leading to the $D+C+R$ coexistence. In contrast to the steady rewarding model, however, here the coexistence is not rooted in a dynamical invasion process of the form $D \to C \to R \to D$, but rather it is due to a static equilibrium. For details concerning the dynamical invasion fronts brought about by steady rewarding we refer to \cite{szolnoki_epl10}, while here we elaborate further on the static equilibrium that is characteristic for adaptive rewarding. Figure~\ref{alliance} (left) features a cross-section of the phase diagram presented in Fig.~\ref{phd_r4_4} at $\Delta=1.5$. It can be observed that as the cost of rewarding ($\alpha$) increases, the pure $R$ phase transform into the three-strategy $D+C+R$ phase, which for still higher values of $\alpha$ becomes the $D+C$ phase. This indicates that as rewarding cooperators loose their ability to deter defectors, they also simultaneously enable the existence of cooperators. Since the value of $r$ is sufficiently high, cooperators can coexist with defectors, in fact forming an alliance with them to compete against rewarding cooperators. The emergence of this alliance can also be inferred from the cross-section plot, where $\rho_C$ and $\rho_D$ change simultaneously as $\alpha$ increases but all the while their ratio remains approximately the same. A characteristic spatial pattern attesting to this fact is presented in Fig.~\ref{alliance} (right), where the $D+C$ patches, which are locally similar to the stable morphology plotted in the upper right panel of Fig.~\ref{indirect}, are surrounded by invading green $R$ players. For the later the cost of rewarding is simply too high to eliminate defectors, which brings along the ``second-order free-riders'' to form the $D+C$ free-riding axis. It is also worth pointing out that as soon as rewarding cooperators die out, the fractions of $D$ and $C$ strategies seizes to vary, indicating that the two indeed form an alliance that depends only on the value of $r$.

If, however, the adaptive rewarding response is made more severe while at the same time rewards remain sufficiently affordable, the three-strategy phase terminates into a defector-free state, denoted as $R(C)$ in Fig.~\ref{phd_r4_4}. The absence of defectors makes cooperators and rewarding cooperators two equivalent strategies. Note that there is a constant drift-towards non-rewarding if defectors fail to spread. This can be either because they are altogether missing, as is the case in the $R(C)$ phase, or because they are not within the immediate neighborhood of $R$ and thus spread undetected. The evolutionary process proceeds without surface tension via logarithmical slow coarsening, as is characteristic for the universality class of the voter model \cite{dornic_prl01}. In \cite{helbing_pre10c}, albeit within a model based on steady punishment, we have demonstrated that the prevalence of ``active cooperators'', here players adopting strategy $R$, can be accelerated very effectively by means of rare mutations. The later give rise to occasional defectors, who in turn mediate the winner similarly as described by the indirect territorial battle in the realm of the $D+C \to D+R$ transition.

\begin{figure}
\centerline{\epsfig{file=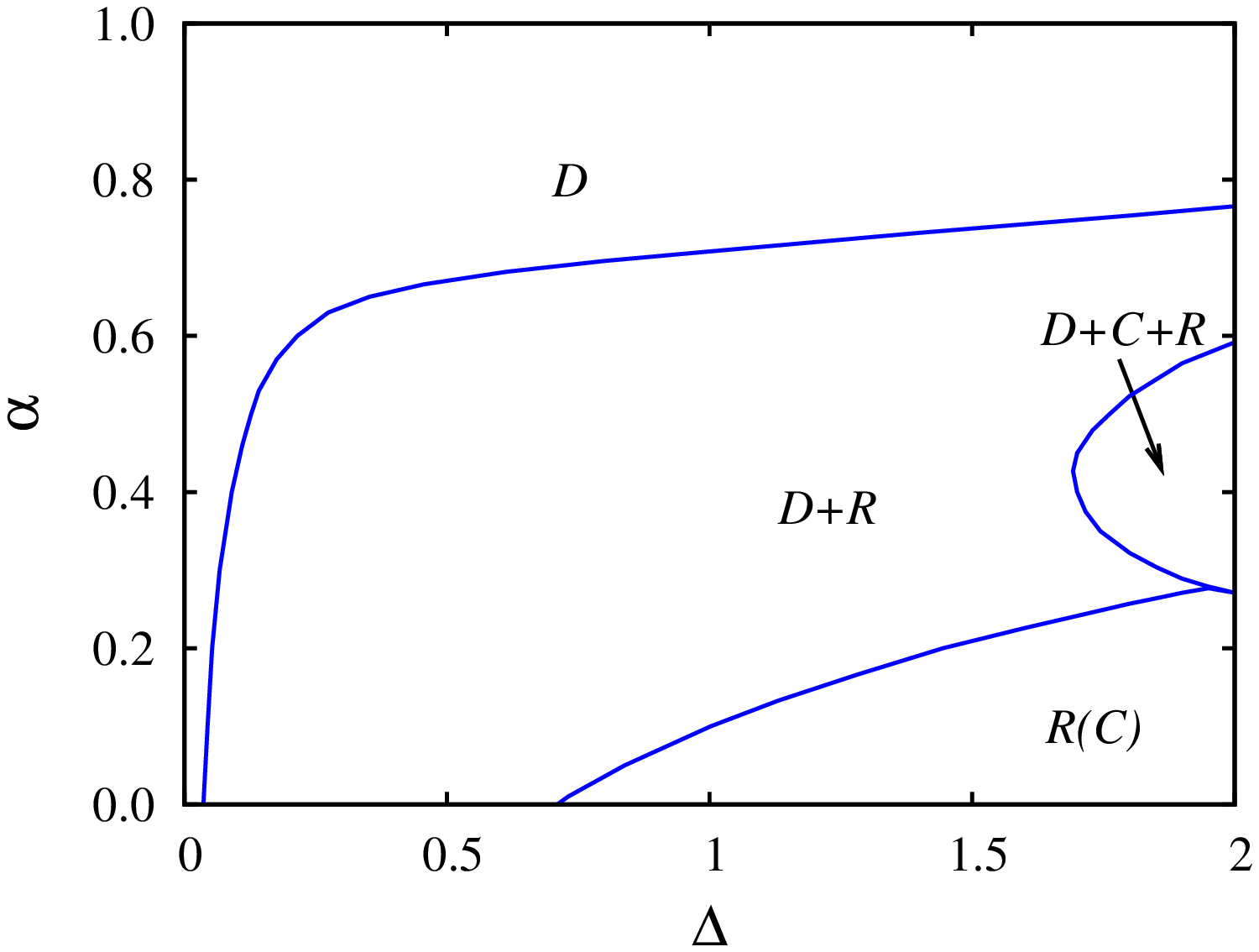,width=8cm}\epsfig{file=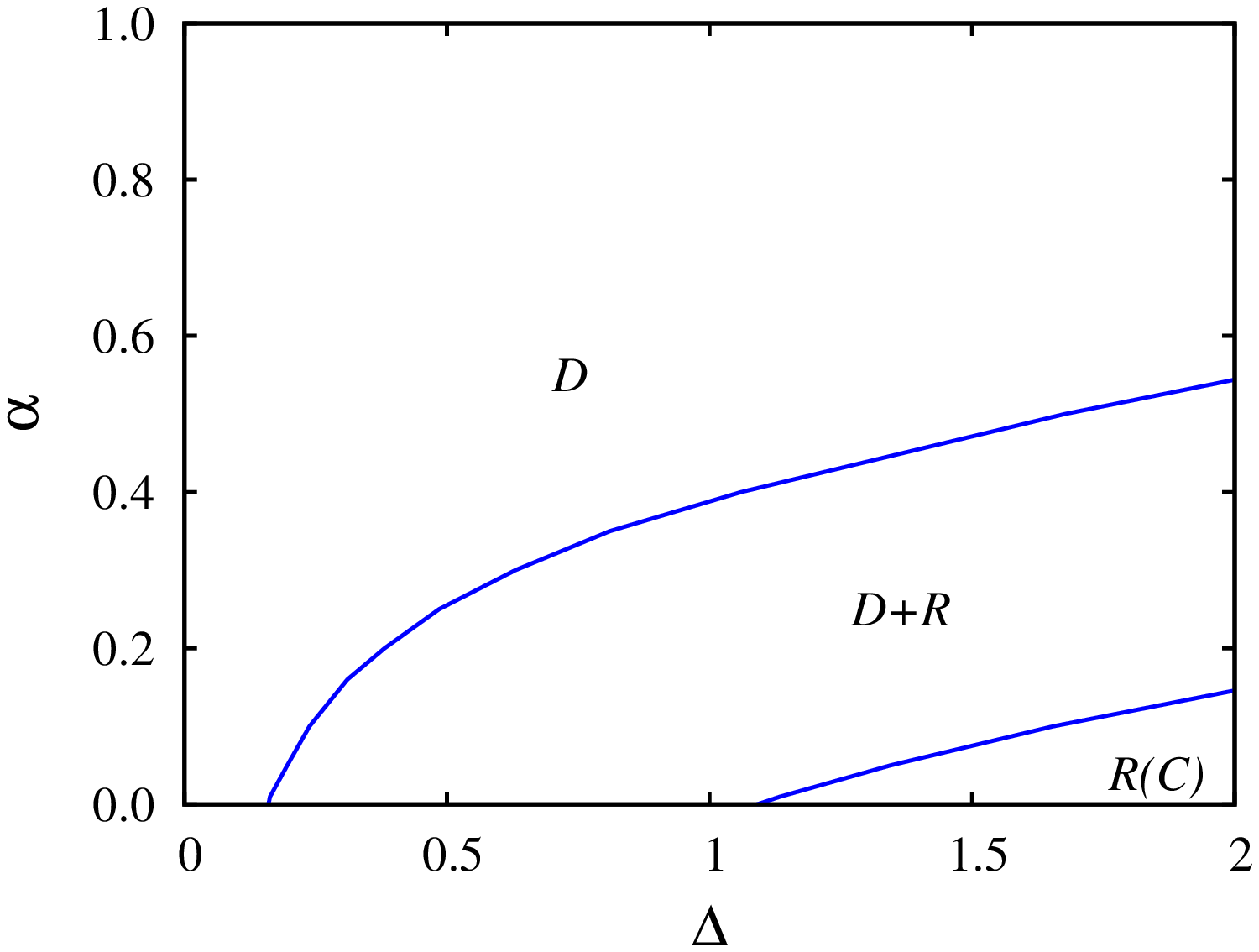,width=8cm}}
\caption{Full $\Delta-\alpha$ phase diagrams, as obtained at $r=3.5$ (left) and $r=2$ (right). As in Fig.~\ref{phd_r4_4}, blue solid lines depict continuous second-order phase transitions and symbols mark the surviving strategies in the stationary state. Since the synergetic effects of collaborative efforts are too weak, cooperators can no longer survive alone in the presence of defectors. Accordingly, the $D+C$ phase is missing. Instead, as $\Delta$ increases, and if $\alpha$ is sufficiently small, the pure $D$ phase gives way to the two-strategy $D+R$ phase, which may further transform into the three-strategy $D+C+R$ phase, but only if $r$ is sufficiently large (left). At $r=2$, for example, the three-strategy phase is no longer attainable on the considered $\Delta-\alpha$ plane. For small rewarding costs the defector-free $R(C)$ phase is obtained (having the same properties as described for $r=4.4$), although its area shrinks continuously as $r$ increases.}
\label{phd_r3_5}
\end{figure}

If the added value of collaborative efforts is smaller, i.e., if $r$ decreases, the phase diagrams change significantly, primarily because the $D+C$ alliance is no longer possible. Figure~\ref{phd_r3_5} features two phase diagrams, left as obtained for $r=3.5$ and right as obtained for $r=2$, where the differences if compared to Fig.~\ref{phd_r4_4} are clearly inferable. If the cost of rewarding is substantial, defectors are the only ones to survive. Naturally, the lower the value of $r$ the lower the value of $\alpha$ that still warrants defector dominance. The pure $D$ phase becomes the two-strategy $D+R$ phase by means of a continuous phase transition even at small $r$, if only the value of $\Delta$ is not too small and the value of $\alpha$ is not too large. Continuing further towards more efficient rewarding may lead to the defector-free $R(C)$ state, which has the same properties as described above for $r=4.4$. As with the $D+R$ phase, the extent of the $R(C)$ region shrinks expectedly with decreasing $r$ towards higher $\Delta$ and lower $\alpha$.

\begin{figure}
\centerline{\epsfig{file=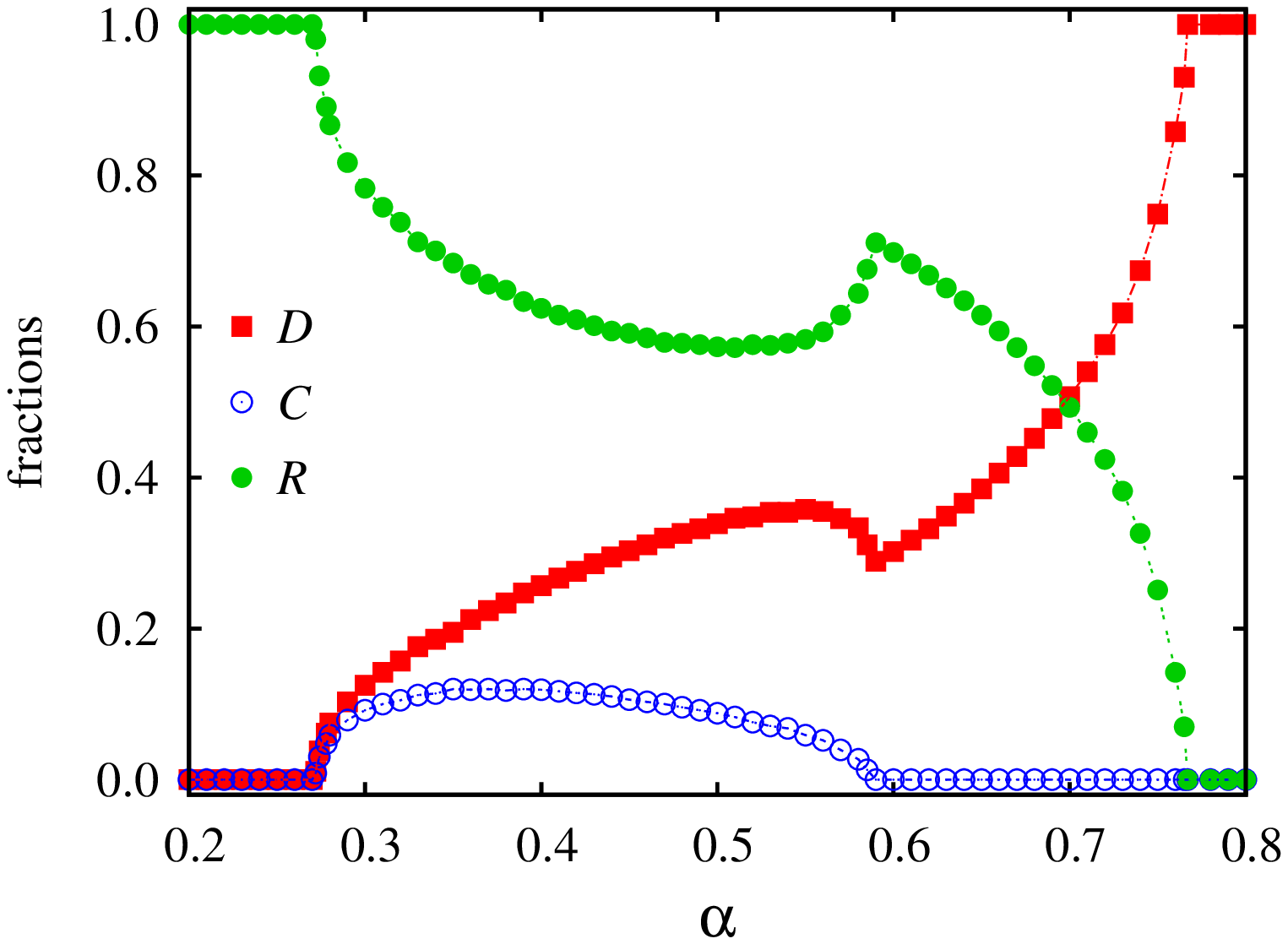,width=8cm}\epsfig{file=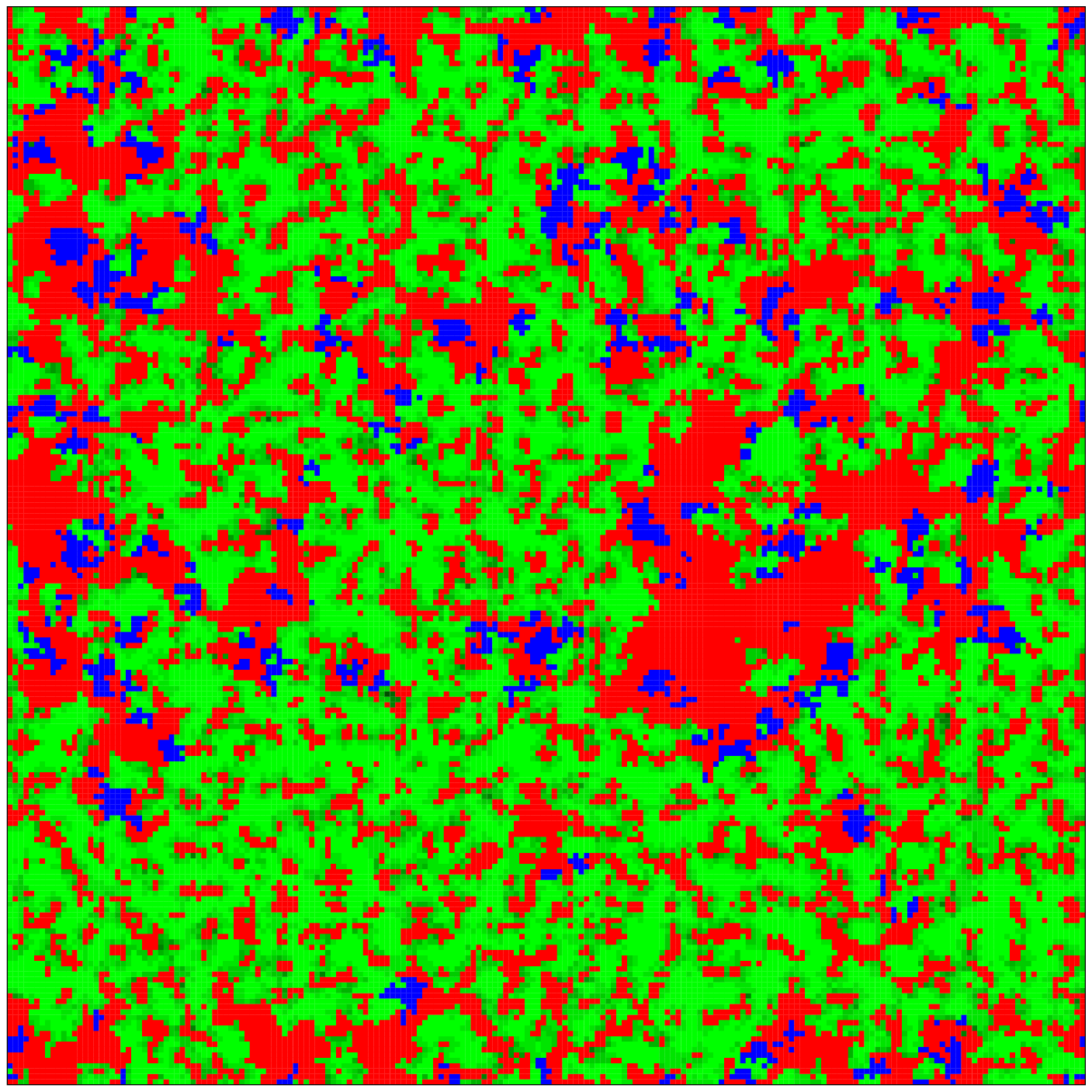,width=6cm}}
\caption{Left panel features a cross section of the phase diagram presented in Fig.~\ref{phd_r3_5} (left), as obtained at $\Delta=2.0$. As $\alpha$ increases the rewarding cooperators first give way to a three-strategy $D+C+R$ phase, but further on persevere longer than ``second-order free-riders''. At smaller values of $r$ the latter require a delicate balance of conditions to survive, and can do so only along the $D+R$ interfaces. Right panel depicts a characteristic snapshot of such a three-strategy phase, which was taken at $\Delta=2.0$ and $\alpha=0.55$. Small and rare patches of cooperators (blue) can survive where defectors (red) and rewarding cooperators (green) meet.}
\label{no_alliance}
\end{figure}

The ``second-order free-riders'', on the other hand, can survive only in the three-strategy $D+C+R$ phase, but its existence is limited to high values of $\Delta$, intermediate $\alpha$ and still sufficiently high values of $r$, as can be observed by comparing the left and right panels of Fig.~\ref{phd_r3_5}. Importantly, this three-strategy phase is qualitatively different from the one described above for $r=4.4$. As mentioned, because of lower $r$ here cooperators cannot survive alone if surrounded solely by defectors. In fact, they can survive only where defectors and rewarding cooperators meet, i.e., along the $D+R$ interfaces. The characteristic snapshot presented in Fig.~\ref{no_alliance} (right) confirms such a spatial configuration within the three-strategy phase. Details of its emergence are inferable from the cross-section of the phase diagram presented in Fig.~\ref{no_alliance} (left), which reveals that as $\alpha$ exceeds a critical value the efficiency of $R$ weakens to the point where defectors are able to survive. The stable presence of a small fraction of cooperators, surviving at the $D+R$ interfaces, accompanies this transition. Interestingly, as $\alpha$ is further increased the first to extinct are not rewarding cooperators but the ``second-order free-riders'', who fail to harvest the benefits of decreased rewarding efficiency. This indicates that, especially at small synergy factors, only a fine balance of all the other parameters enables the survival of ``second-order free-riding''.

\subsection{Reward versus punishment}
\label{vs}

\begin{figure}
\centerline{\epsfig{file=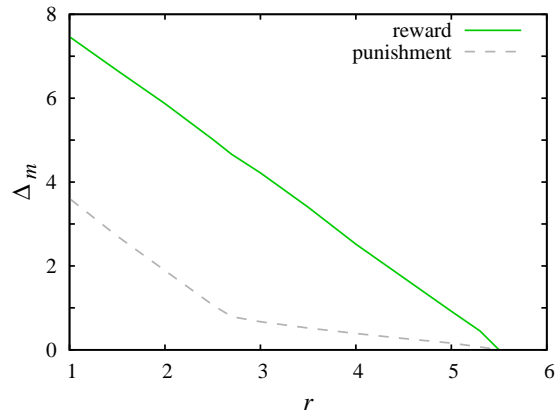,width=8cm}}
\caption{Minimally required strength of the adaptive response $\Delta_m$ that warrants extinction of defectors in dependence on the synergy factor $r$ at $\alpha=0.4$. Compared is the efficiency of adaptive rewarding (solid green line) with that of adaptive punishment (dashed gray line), and it can be observed clearly that the latter is more effective.}
\label{punish}
\end{figure}

Finally, we address the ``stick versus carrot'' dilemma within the realm of adaptive modeling. To do so, we first focus on the competition solely between defectors and rewarding cooperators. The question is, given a fixed cost of administrating rewards $\alpha$, what is the minimally required value of $\Delta$ that warrants the complete elimination of defectors? The answer is presented in Fig.~\ref{punish} as a function of the synergy factor $r$ (solid green line). Next, we answer the same question again, but replacing the rewarding cooperators with punishing cooperators. For consistency we use the same value of $\alpha$, but accordingly it now represents the punishment cost rather than the cost of rewarding. The dashed gray line in Fig.~\ref{punish}, depicting the results for adaptive punishment, falls significantly below the one obtained with adaptive rewarding. This leads to the conclusion that adaptive punishment, which we studied separately in \cite{perc_njp12}, is more effective than adaptive rewarding in warranting defector-free states.

\begin{figure}
\centerline{\epsfig{file=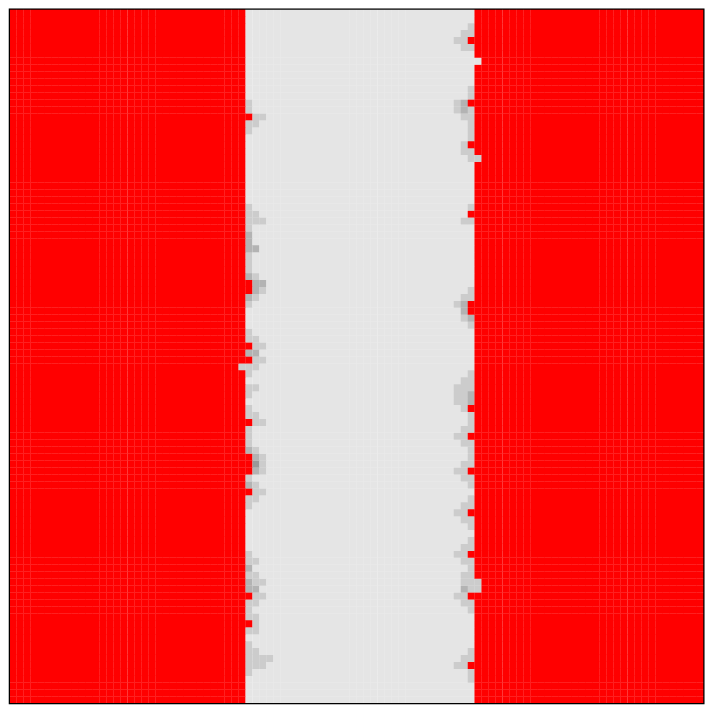,width=2.7cm}\epsfig{file=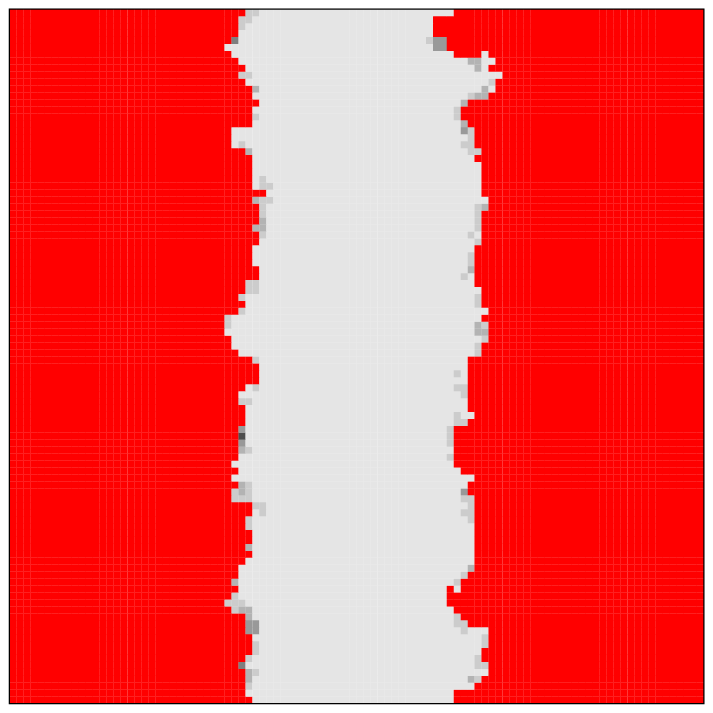,width=2.7cm}\epsfig{file=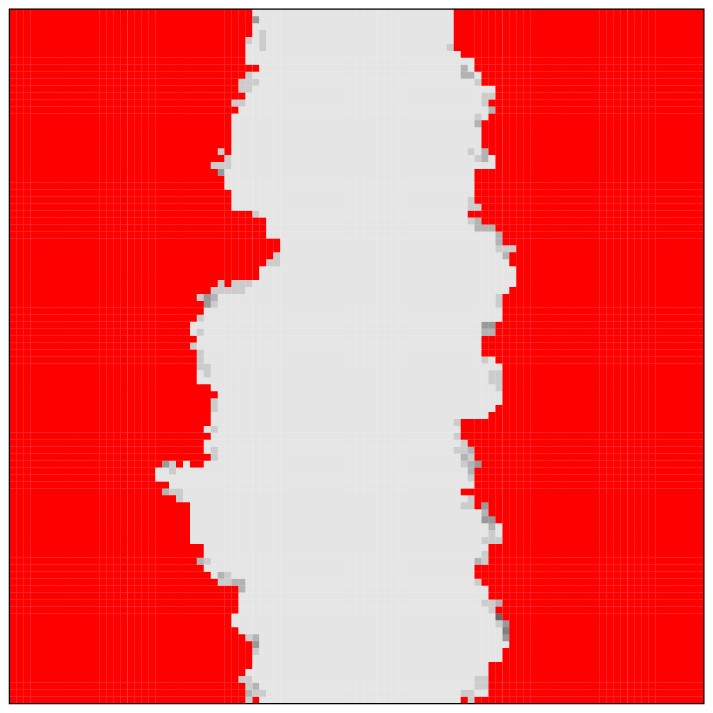,width=2.7cm}\epsfig{file=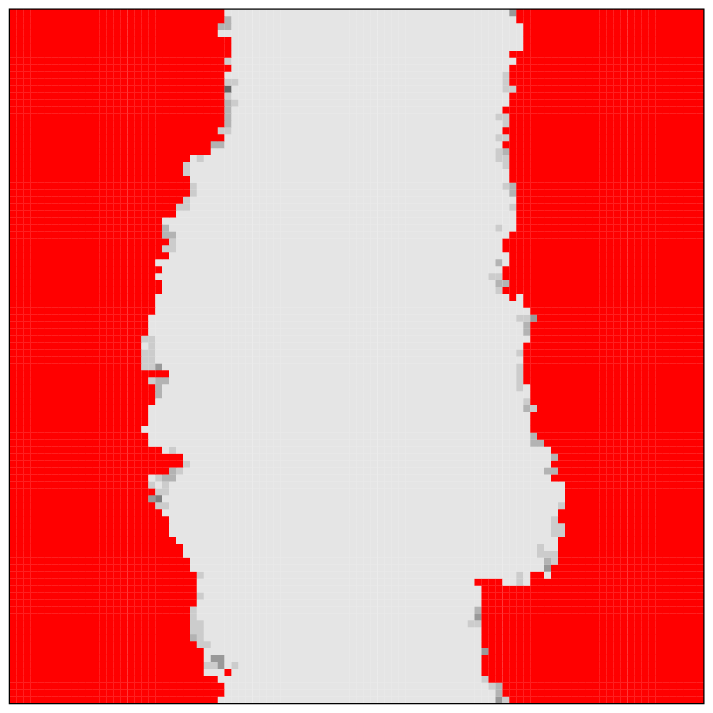,width=2.7cm}\epsfig{file=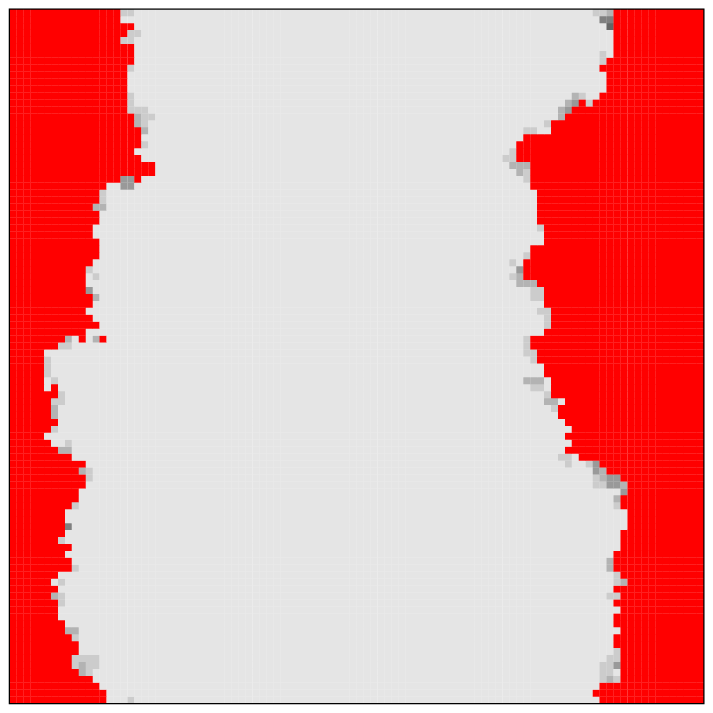,width=2.7cm}\epsfig{file=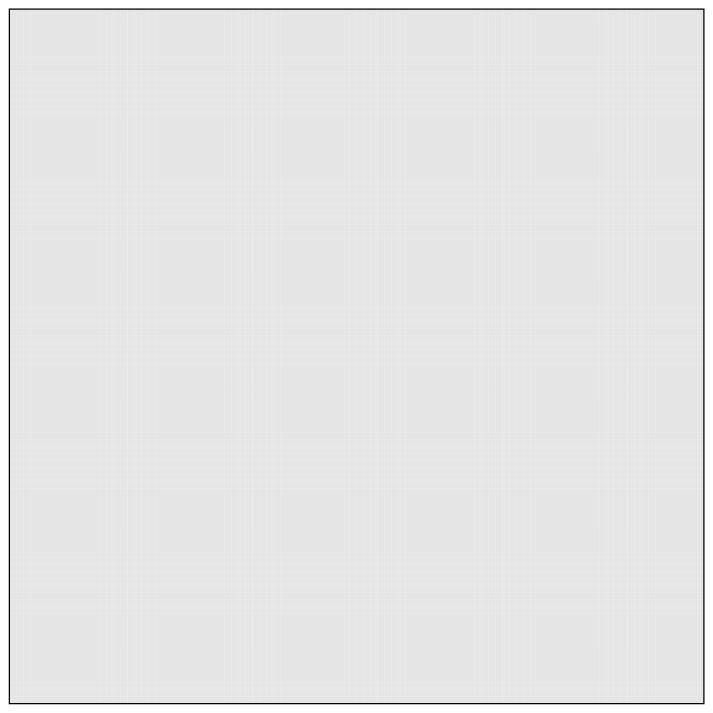,width=2.7cm}}
\centerline{\epsfig{file=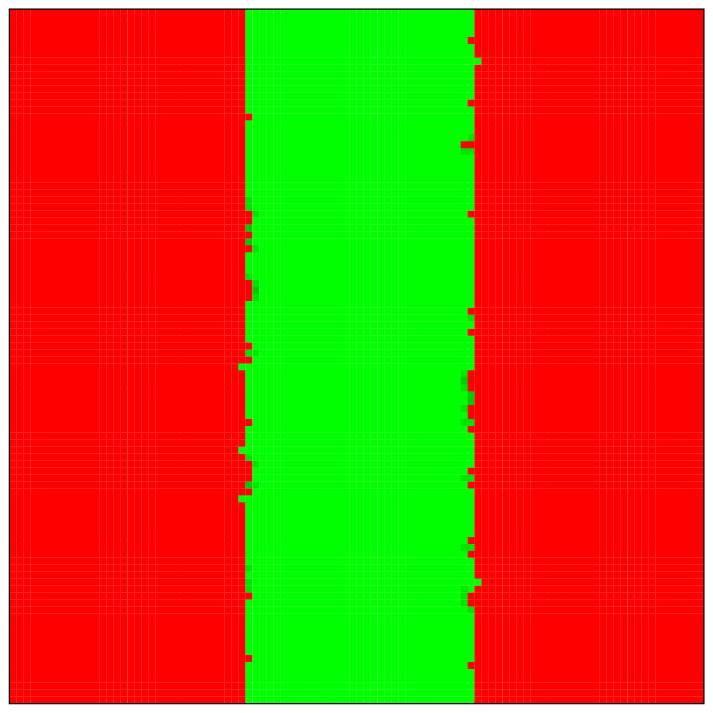,width=2.7cm}\epsfig{file=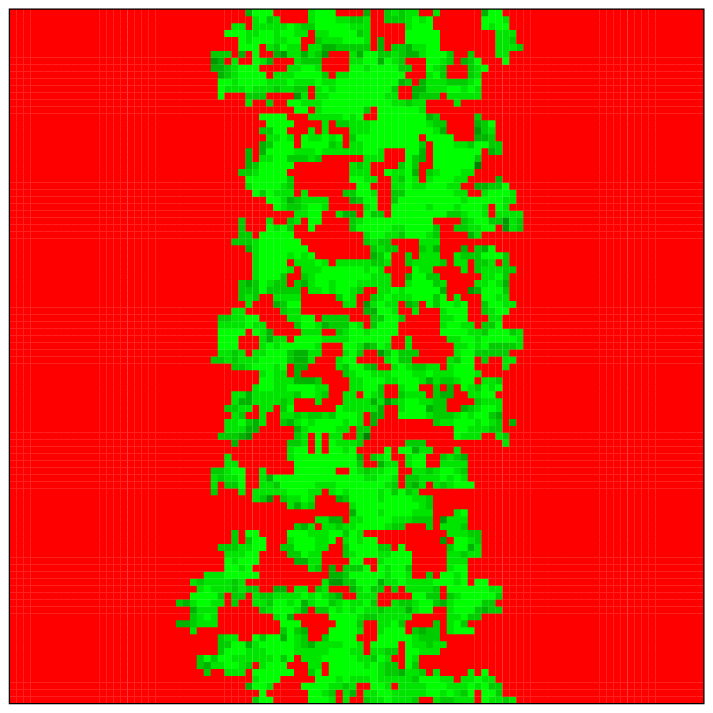,width=2.7cm}\epsfig{file=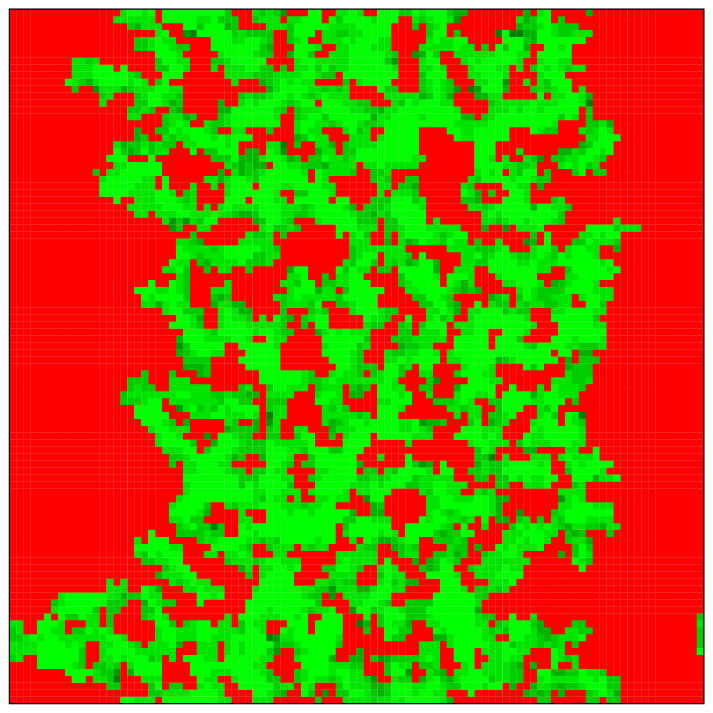,width=2.7cm}\epsfig{file=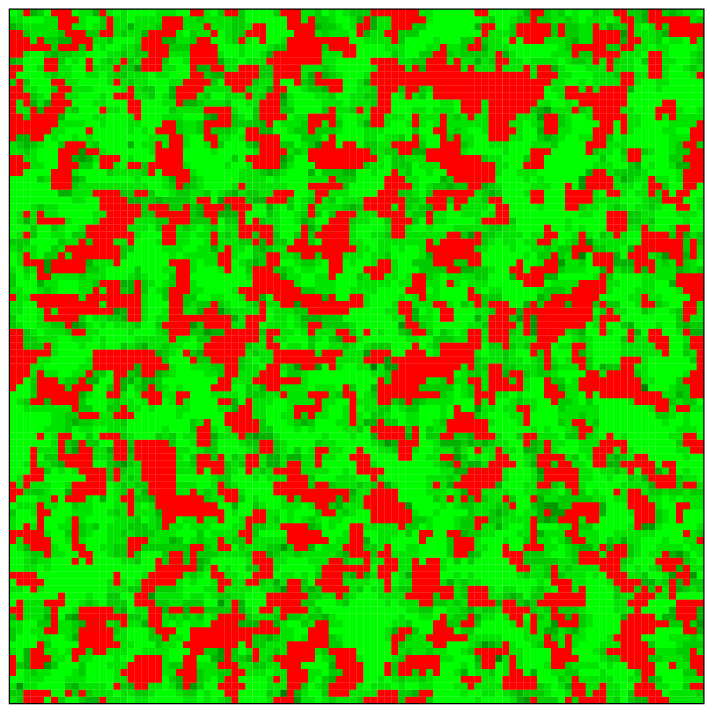,width=2.7cm}\epsfig{file=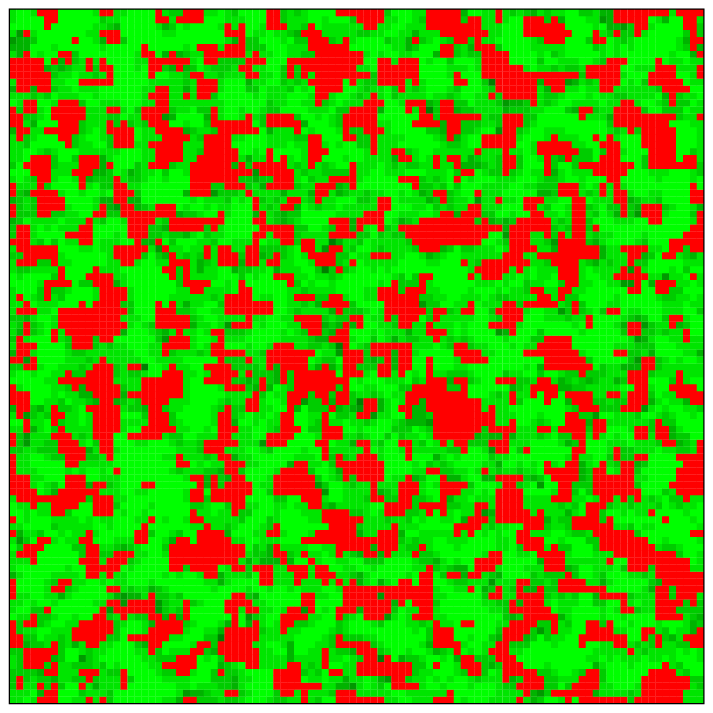,width=2.7cm}\epsfig{file=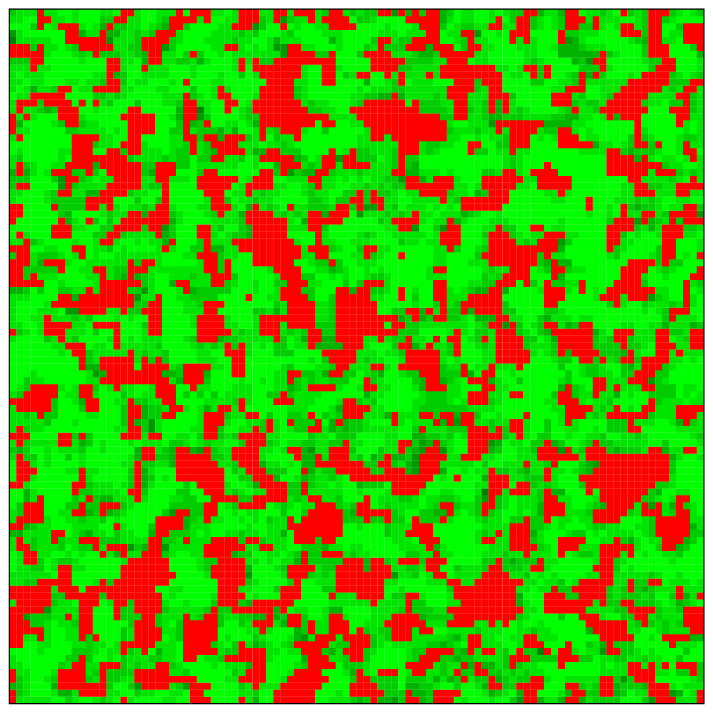,width=2.7cm}}
\caption{Comparison of the evolution of interfaces separating punishing cooperators and defectors (top row), and rewarding cooperators and defectors (bottom row). It can be observed that while rewarding cooperators (green) advance faster into the territory of defectors (red), the punishing cooperators (gray) are relentlessly bent on keeping their phase compact. Although therefore advancing less fast, they ultimately succeed in completely eliminating the defectors. Rewarding cooperators, on the other hand, have to make do with their coexistence. Note that darker shades of gray (green) denote players with higher punishing (rewarding) activity. The parameter values are the same for both cases, namely $r=2$, $\Delta=2$ and $\alpha=0.4$, while the snapshots were taken at $1$, $70$, $300$, $1000$, $3000$ and $6000$ full Mote Carlo Steps.}
\label{compare}
\end{figure}

An intuitive explanation as to why this is the case is presented in Fig.~\ref{compare}, where we follow the evolution of interfaces separating defectors and punishing cooperators (top row) as well as defectors and rewarding cooperators (bottom row) under identical conditions. It can be observed that while rewarding cooperators are more successful in penetrating the area of defectors, the punishing cooperators advance less fast but maintain a compact phase. For example, in the third snapshot from the left, some rewarding cooperators have already reached the border of the lattice while punishing cooperators have yet to advance notably. However, rewarding cooperators have to pay a price for their over-aggressive invasion, namely an irregular interface that facilitates the coexistence with defectors. Paradoxically, the less aggressive effect of punishment, which focuses on repairing the cracks in the phalanx rather than on advancing into the territory of defectors at any cost, turns out to be more effective at the end. Punishing cooperators rise to complete dominance with the aid of a near flawless support of network reciprocity \cite{nowak_n92b}. Rewarding cooperators, on the other hand, sacrifice the latter for a faster advancement, but therefore fail to create the desired defector-free state. The Achilles heel of rewarding is thus an excessively aggressive invasion of defectors that neglects the benefits of network reciprocity.

\section{Summary}
\label{summary}
We have shown that adaptive rewarding creates several evolutionary advantages by means of which public cooperation is promoted, in particularly many such that go beyond those warranted by steady rewarding \cite{szolnoki_epl10}. Phase diagrams and the corresponding analysis of spatial patterns reveal that, if the added value of collaborative efforts is substantial, rewarding cooperators fight an indirect territorial battle with the cooperators. The catalysts are the defectors, who essentially determine the victor depending on who can invade them more successfully. If the parameters determining adaptive rewarding are set properly, most notably if the rewards are sufficiently yet not too cheap and the response to invading defectors is sufficiently strong, the three competing strategies form a stable phase wherein defectors form a free-riding alliance with cooperators, i.e., ``second-order free-riders'', to compete against rewarding cooperators. This three-strategy phase can also be observed at intermediate multiplication factors, although its extent in the phase diagrams shrinks continuously as the synergetic effects of cooperation are lowered, and accordingly the $D+C$ alliance becomes more and more difficult. It is also worth emphasizing that the spatial dynamics enabling the three-strategy phase changes. While for sufficiently high multiplication factors cooperators can survive alone in the presence of defectors, at lower values of $r$ they can avoid extinction only in the immediate vicinity of $D-R$ interfaces. If either the cost of rewarding is decreased further or the adaptive response is made even more severe, the coexistence is terminated, which leads to a defector-free state. Due to a constant drift towards non-rewarding in the absence of defectors that could spread successfully, cooperators and rewarding cooperators become equivalent strategies, and accordingly the victor is determined via slow logarithmic coarsening, as known from the voter model \cite{dornic_prl01}. In the majority of cases, however, rewarding cooperators occupy the larger portion of the square lattice at the time defectors die out, and accordingly they are the more likely winners. This competition becomes even more biased in the presence of rare mutations.

Comparing the outcome with the one elicited by adaptive punishment \cite{perc_njp12}, we find that the supreme efficiency of rewarding cooperators in terms of invading defectors lessens the effectiveness of network reciprocity, which in turn designates the slower advancing adaptive punishers as the more effective and indeed the more successful strategy. We report that the minimally required fine to reach a defector-free state is much lower than the minimal reward needed to achieve the same goal. Thus, while a deep invasion of isolated players into the territory of defectors is better supported by rewarding, punishers can reach the collective target of eliminating defectors only by collaborating and ``holding the line''. The later statement is reminiscent of an instruction that is frequently given to soldiers engaging in combat, highlighting the continued importance of network reciprocity despite additional, and locally more effective means to overcome defection. The uncovered Achilles heel of rewards may also provide further clues as to why order and justice in the society are maintained by law that focuses on sanctioning rather than rewarding -- although the former acts more subtle and requires a higher coherence between group members, at the same conditions it provides a higher collective well-being for the whole community.

\ack
This research was supported by the Hungarian National Research Fund (grant K-101490) and the Slovenian Research Agency (grant J1-4055).

\section*{References}
\providecommand{\newblock}{}

\end{document}